\documentclass[%
 aip,
cp,  
 amsmath,amssymb,
 reprint,%
]{revtex4-2}

\usepackage{graphicx}
\usepackage{dcolumn}
\usepackage{bm}

\usepackage[utf8]{inputenc}
\usepackage[T1]{fontenc}
\usepackage{mathptmx} 

\begin{document}

\title{New Results About The Revolutionary Bolometer Assembly Of BINGO}

\author{A. Armatol on behalf of the BINGO collaboration} 
 \email{antoine.armatol@cea.fr}
\affiliation{IRFU, CEA, Universit\'e Paris-Saclay, Saclay, France}
%

\date{\today} 

\begin{abstract}
Searching for neutrinoless double-beta decay (0$\nu$2$\beta$) is one of the main experimental challenges of modern physics since its discovery would prove the Majorana nature of neutrinos. One experimental technique is given by cryogenic detectors named bolometers that are really promising for this purpose. The current generation tonne-scale experiment CUORE using this technology is putting the best limit on $^{130}$Te 0$\nu$2$\beta$ half-life with TeO$_2$ crystals but its sensitivity is limited by its background. Therefore, it will be followed by the next generation experiment CUPID (CUORE Upgrade with Particle IDentification) that will study $^{100}$Mo embedded inside Li$_2$MoO$_4$ crystals in order to reduce the $\gamma$ background. It will also read the scintillation light produced by Li$_2$MoO$_4$ by adding another Ge bolometer acting as a light detector next to the main absorber to reject the $\alpha$ background. Thanks to that, CUPID will reach a sensitivity 2 orders of magnitude higher than CUORE. However, in the case where this is not enough to detect 0$\nu$2$\beta$, BINGO (Bi-Isotope Next Generation 0$\nu$2$\beta$ Observatory) is preparing the next-next generation of bolometric experiments. To improve the 0$\nu$2$\beta$ discovery sensitivity, the goal is to reduce drastically the number of background events in the region of interest and to combine the use of the two previously cited isotopes: $^{130}$Te and $^{100}$Mo. To achieve this goal, BINGO is proposing to implement an active cryogenic veto to suppress the external $\gamma$ background, to use Neganov-Trofimov-Luke effect to increase light detector sensitivity and to use a revolutionary detector assembly to  reduce the total surface radioactivity contribution. In this article, we will focus on the latter and present the latest results obtained with two 45$\times$45$\times$45 mm$^3$ Li$_2$MoO$_4$ crystals. This is the first time that such a way to assemble bolometers of this size is tested.

\end{abstract}

\maketitle

\section{\label{sec:INTRO}Introduction}

One of the big questions of neutrino physics is its nature: is it a Dirac particle, like the other particles of the Standard Model, or a so-called Majorana particle (i.e its particle coincides with its antiparticle)? The only way to answer experimentally to this question would be to observe a process possible only in the latter case: the neutrinoless double beta decay (0$\nu$2$\beta$) \cite{Deppisch_2012,ReviewNDBD}. It is a hypothetical alternative mode of the two-neutrino double beta decay (2$\nu$2$\beta$) \cite{DBD}, already observed for 11 nuclei, but where only two electrons are emitted in the final state. This observation would then prove also the violation of lepton number conservation and therefore have a lot of consequences in particle physics \cite{Asym1}. The experimental signature is a peak at the $Q$-value of 0$\nu$2$\beta$ in the energy spectrum of the two emitted electrons. Since this decay is extremely rare, experimental challenges are numerous. For example, the experiment should have a huge sensitive exposure (i.e. time of exposure $\times$ mass of 0$\nu$2$\beta$ candidate isotope), ideally zero background events in the region of interest (ROI), an excellent energy resolution and a high detection efficiency. 

One of the most promising detectors for this search are bolometers. They are cryogenic detectors operated at around 20 mK composed of a main absorber that embeds a 0$\nu$2$\beta$ candidate. The principle is rather simple: a small thermistor is glued on the absorber surface in order to measure the rise of temperature produced by the energy release of a particle inside it. They are gathering all the important features previously listed, the only remaining point is the background that does not depend directly on the detection method but more on materials and technologies used to reject it. It is possible to quantify this background with a parameter called the background index b that represents the amount of expected background events in the ROI in counts per kilogram, per keV, per year (ckky). 

The first tonne-scale cryogenic bolometer array, CUORE, is currently in data taking period in the underground laboratory of Gran Sasso (LNGS) in Italy. It observes $^{130}$Te embedded in TeO$_2$ crystals. It has recently achieved a 1 ton.yr exposure and has put the best limit on $^{130}$Te 0$\nu$2$\beta$ half life (T$^{0\nu}_{1/2}$>2.2$\times$10$^{25}$ yr) \cite{CUORE2022}. However, CUORE is not background free and shows a background index of $b\simeq$10$^{-2}$ ckky, dominated by $\alpha$ particles coming from the contamination of surrounding materials \cite{CUOREbckg}. In the future, CUORE will have a successor named CUPID \cite{CUPIDCDR} for which the goal is to increase the sensitivity to the half life to a value superior to 10$^{27}$ yr and so to reduce the background index by two orders of magnitude and reach $b\simeq$10$^{-4}$ ckky. To achieve this objective, CUPID will switch of isotope using $^{100}$Mo, taking benefit of its greater Q-value (3034 keV instead of 2527 keV for $^{130}$Te). The isotope of interest will be embedded inside Li$_2$MoO$_4$ crystals which are scintillators. Hence, by putting an auxiliary bolometer made of germanium acting as a light detector next to the main absorber, each event will produce a heat and a light signal. This feature allows to discriminate $\beta/\gamma$ events from $\alpha$ since the latter are producing a different amount of light when interacting in the crystal \cite{SCINTILLATION}. A rejection of more than 99\% of the $\alpha$ events in addition to the good performance of Li$_2$MoO$_4$ crystals have been proven by the CUPID-Mo experiment \cite{CMO,CUPID_Mo}, validating this technology for CUPID. 

At this point, if 0$\nu$2$\beta$ is still eluding us, it means that the sensitivity to the half life of next-next generation experiments will have to be pushed even further. This implies to increase the total sensitive exposure but also to reduce the amount of background in the ROI. The BINGO project takes place in this framework: starting from CUPID technology, it proposes innovative methods to reduce drastically the number of background events in the ROI for bolometric experiments and to prepare this next-next generation. The main objective is to show that, with the right exposure, a background index of $b\simeq$10$^{-5}$ ckky is reachable with BINGO improvements, doubling the current CUPID expected sensitivity.

\section{\label{sec:IMPRO}BINGO IMPROVEMENTS}

The first main innovation proposed by BINGO is to combine the use of two of the most promising 0$\nu$2$\beta$ candidates: $^{100}$Mo embedded inside Li$_2$MoO$_4$ crystals and $^{130}$Te embedded in TeO$_2$ crystals. The use of $^{130}$Te is interesting since its natural abundance is really high (34\%) making the enrichment of the crystals not mandatory. Therefore, combine it with $^{100}$Mo allows to reach larger and larger masses at a reasonable cost. It implies also some challenges since the expected 0$\nu$2$\beta$ peak of $^{130}$Te is expected in a region with a rather high amount of background $\gamma$ events (its $Q$-value is below the natural $\gamma$ radioactivity endpoint). Hence, a way to reject these background events has to be implemented. Moreover, to use TeO$_2$ crystals as luminescent bolometers, BINGO must have light detectors with a sensitivity high enough to detect the Cerenkov light emitted by the $\beta/\gamma$ events. Indeed, these crystals are mainly emitting light only through this process \cite{Cerenkov_TeO}.

BINGO will also implement for the first time a cryogenic active veto around the detector area. It will be made of scintillators (BGO or ZnWO) and thanks to coincidence studies, it will be possible to reject the external $\gamma$ background, essential to use the TeO$_2$ crystals in an effective way. A first comparison between the two candidates is described in Ref~\onlinecite{BINGO1}. The scintillation light of this veto will be read by Ge bolometers that should have a high detection sensitivity to detect as much as possible $\gamma$'s crossing the veto.

In order to boost the light signal, BINGO will take benefit of the so-called Neganov-Trofimov-Luke (NTL) effect on its light detectors \cite{NLffect}. Aluminium electrodes will be evaporated on the Ge wafer surface allowing to apply a voltage difference between them. When charges are created by an event, they are then drifted to the electrodes, amplifying the heat produced inside the absorber and thus the signal. These NTL light detectors allow on the other hand to reach a sensitivity high enough to detect the Cerenkov light of TeO$_2$ crystals and to use them, like Li$_2$MoO$_4$, as scintillating bolometers \cite{NLTeO}.

BINGO is also studying a revolutionary way to assemble bolometers. A single module is composed of two crystals with a copper piece placed in the middle. The crystals are kept attached to the copper thanks to nylon wires. Moreover, light detectors are placed in a vertical position in between the crystal and the holder to completely shield the absorber from the copper surface radioactivity. In this configuration, the amount of passive surrounding material is reduced by 2 orders of magnitude compared to the CUORE configuration. It is geometrically minimizing the contribution of surface radioactivity to the background. In the rest of this paper we will present the new results we have obtained with this assembly using 45$\times$45$\times$45 mm$^3$ Li$_2$MoO$_4$ crystals.

All these technologies should allow to reach the objective of 10$^{-5}$ ckky for the background index and push the sensitivity of bolometric experiments to a new level.

\subsection{\label{sec:level2}Results of the nylon assembly with 45$\times$45$\times$45 mm$^3$ Li$_2$MoO$_4$ crystals}

\begin{figure}[h]
\includegraphics[width=0.4\linewidth]{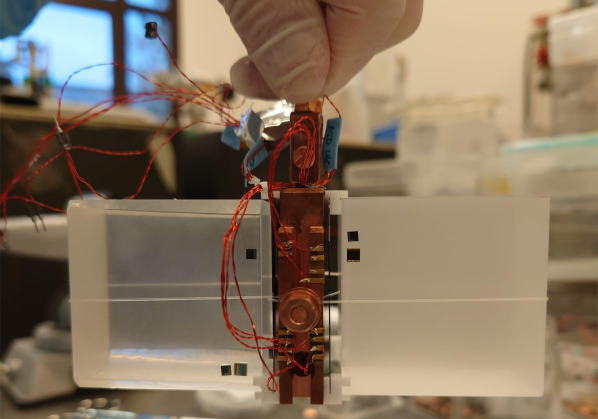}
\caption{\label{fig:epsart} The nylon wire assembly. On the left is LMO56b and on the right is LMO21-2. One light detector per crystal was placed vertically between it and the holder.}
\end{figure}

This way to assemble the detectors was first tested inside a pulse-tube cryostat with one module composed of two 20$\times$20$\times$20 mm$^3$ Li$_2$MoO$_4$ crystals. The goal was to make sure that the nylon wires were keeping the crystals well fixed even at low temperature, without preventing them to obtain satisfying bolometric performance. The results of this test and a better description of the assembly can be found in Ref~\onlinecite{BINGO1}. Given the good obtained results, we decided to move on with a bigger crystal size of 45$\times$45$\times$45 mm$^3$ which is the typical size foreseen for CUPID crystals. 
We improved the design of the holder between the two tests, reducing even more the amount of copper. Moreover, a tool has been developed to help to put in place the nylon wires, making the assembly easy and straightforward. The thermistors glued on the crystals were NTD (Neutron Transmutation Doped) Ge \cite{NTDGe} for which the resistance is depending on the temperature. The module (Fig.~\ref{fig:epsart}) was then put inside the same above ground pulse-tube cryostat as for the test with the small crystals. Several measurements were acquired at different cryogenic temperatures, each time the optimum bias to read the NTDs resistance was determined with load curves \cite{khalife}.

\begin{table}[h]
\caption[]{Best performance at 15 mK of the detectors }
\label{tab:resultsnylon}
\begin{center}
\begin{tabular}{|c|c|c|c|c|}
\hline
Detector & $R_{NTD}$ (M$\Omega$)  & Sensitivity (nV/keV)  & FWHM$_{noise}$ (keV) & Light yield (keV/MeV) \\
\hline
LMO21-2 & 8.3 & 58 & 4.9 & 0.15  \\
\hline
LMO56b & 41.2 & 66 & 13 & 0.20  \\
\hline
\end{tabular}
\end{center}
\end{table}

\begin{figure}[]
\begin{minipage}{0.5\linewidth}
\centerline{\includegraphics[width=0.8\linewidth]{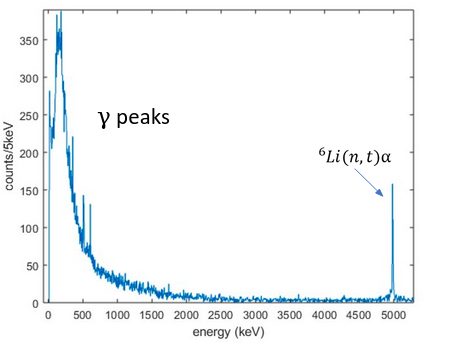}}
\end{minipage}
\hfill
\begin{minipage}{0.49\linewidth}
\centerline{\includegraphics[width=1\linewidth]{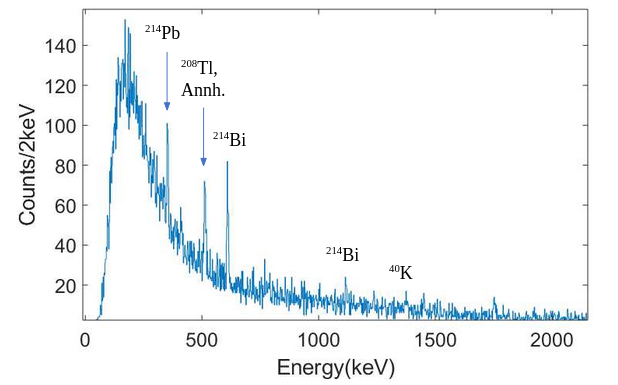}}
\end{minipage}
\caption[]{Energy spectrum obtained for LMO21-2 during the background measurement at 15 mK (left). Same spectrum zoomed in the $\gamma$ region (right).}
\label{fig:nylon}
\end{figure}

We present here the best results obtained for the crystals at 15 mK during a background measurement. The two crystals were called LMO21-2 and LMO56b. The former was measured for the first time while the latter had been already measured in a standard assembly closer to the CUORE one. It was then acting as a reference to check if the nylon assembly was affecting its behaviour. The best performance are shown in Table~\ref{tab:resultsnylon}. LMO21-2 has exhibited a good sensitivity for a crystal of this size and a noise FWHM below 5 keV. The energy resolution measured was also encouraging with 5.95 keV FWHM at 609 keV for the $^{214}$Bi $\gamma$ peak and 14.6 keV FWHM for the $^6$Li neutron capture (Fig.~\ref{fig:nylon}). Concerning LMO56b, the noise level was higher. This can be explained by the fact that an important stress was applied on the NTD by the gluing, elucidating also its high resistance. Moreover, the glue type used to glue the NTD of each crystal are different: for LMO21-2 Araldite epoxy glue was used while UV glue was used for LMO56b. Anyway, this behaviour is independent of the assembly since similar results were observed in the previous measurement inside the standard holder. Hence, both crystals have obtained results validating the nylon wire assembly suitability for bolometric experiments. 

\section{Conclusion}

In view of the BINGO project and its goal to reduce drastically the number of background events in bolometric experiments, we have pursued the test on the revolutionary detector assembly using 45x45x45 mm$^3$ Li$_2$MoO$_4$ crystals. We have obtained results within the requirements of BINGO and have shown that this way to hold crystals could be a solution for next-next generation bolometric experiments. The next step will be to test two modules stacked in a tower in the underground laboratory of Canfranc to set the final performance of this assembly. Moreover, the R\&D is also on-going on the other improvements listed in this paper. At the end, this work will lead to the construction of a demonstrator for BINGO technology in the underground laboratory of Modane: MINI-BINGO. It will be composed of one tower of Li$_2$MoO$_4$ and one of TeO$_2$ with 12 crystals each. It will include all the BINGO improvements such as the nylon assembly, the cryogenic veto and the NTL light detectors. This demonstrator will prove that with the right sensitive exposure, the background index of $b\simeq$10$^{-5}$ ckky is reachable and then set the foundation for the next-next generation of bolometric experiments searching for 0$\nu$2$\beta$.

\begin{acknowledgments}
The BINGO project has received funding from the European Research Council (ERC) under the European Union’s Horizon 2020 research and innovation program (grant agreement No 865844). The INR NASU group was supported in part by the National Research Foundation of Ukraine Grant no. 2020.02/0011. We would like to thank also Ph.~Forget and the mechanical workshop of SPEC for their valuable help.
\end{acknowledgments}

\nocite{*}
\bibliography{aipsamp}

\providecommand{\noopsort}[1]{}\providecommand{\singleletter}[1]{#1}%
\begin{thebibliography}{16}%
\makeatletter
\providecommand \@ifxundefined [1]{%
 \@ifx{#1\undefined}
}%
\providecommand \@ifnum [1]{%
 \ifnum #1\expandafter \@firstoftwo
 \else \expandafter \@secondoftwo
 \fi
}%
\providecommand \@ifx [1]{%
 \ifx #1\expandafter \@firstoftwo
 \else \expandafter \@secondoftwo
 \fi
}%
\providecommand \natexlab [1]{#1}%
\providecommand \enquote  [1]{``#1''}%
\providecommand \bibnamefont  [1]{#1}%
\providecommand \bibfnamefont [1]{#1}%
\providecommand \citenamefont [1]{#1}%
\providecommand \href@noop [0]{\@secondoftwo}%
\providecommand \href [0]{\begingroup \@sanitize@url \@href}%
\providecommand \@href[1]{\@@startlink{#1}\@@href}%
\providecommand \@@href[1]{\endgroup#1\@@endlink}%
\providecommand \@sanitize@url [0]{\catcode `\\12\catcode `\$12\catcode
  `\&12\catcode `\#12\catcode `\^12\catcode `\_12\catcode `\%12\relax}%
\providecommand \@@startlink[1]{}%
\providecommand \@@endlink[0]{}%
\providecommand \url  [0]{\begingroup\@sanitize@url \@url }%
\providecommand \@url [1]{\endgroup\@href {#1}{\urlprefix }}%
\providecommand \urlprefix  [0]{URL }%
\providecommand \Eprint [0]{\href }%
\providecommand \doibase [0]{http://dx.doi.org/}%
\providecommand \selectlanguage [0]{\@gobble}%
\providecommand \bibinfo  [0]{\@secondoftwo}%
\providecommand \bibfield  [0]{\@secondoftwo}%
\providecommand \translation [1]{[#1]}%
\providecommand \BibitemOpen [0]{}%
\providecommand \bibitemStop [0]{}%
\providecommand \bibitemNoStop [0]{.\EOS\space}%
\providecommand \EOS [0]{\spacefactor3000\relax}%
\providecommand \BibitemShut  [1]{\csname bibitem#1\endcsname}%
\let\auto@bib@innerbib\@empty
\bibitem [{\citenamefont {Deppisch}, \citenamefont {Hirsch},\ and\
  \citenamefont {Päs}(2012)}]{Deppisch_2012}%
  \BibitemOpen
  \bibfield  {author} {\bibinfo {author} {\bibfnamefont {F.~F.}\ \bibnamefont
  {Deppisch}}, \bibinfo {author} {\bibfnamefont {M.}~\bibnamefont {Hirsch}}, \
  and\ \bibinfo {author} {\bibfnamefont {H.}~\bibnamefont {Päs}},\ }\bibfield
  {title} {\enquote {\bibinfo {title} {Neutrinoless double-beta decay and
  physics beyond the standard model},}\ }\href {\doibase
  10.1088/0954-3899/39/12/124007} {\bibfield  {journal} {\bibinfo  {journal}
  {Journal of Physics G: Nuclear and Particle Physics}\ }\textbf {\bibinfo
  {volume} {39}},\ \bibinfo {pages} {124007} (\bibinfo {year}
  {2012})}\BibitemShut {NoStop}%
\bibitem [{\citenamefont {Dolinski}, \citenamefont {Poon},\ and\ \citenamefont
  {Rodejohann}(2019)}]{ReviewNDBD}%
  \BibitemOpen
  \bibfield  {author} {\bibinfo {author} {\bibfnamefont {M.~J.}\ \bibnamefont
  {Dolinski}}, \bibinfo {author} {\bibfnamefont {A.~W.}\ \bibnamefont {Poon}},
  \ and\ \bibinfo {author} {\bibfnamefont {W.}~\bibnamefont {Rodejohann}},\
  }\bibfield  {title} {\enquote {\bibinfo {title} {Neutrinoless double-beta
  decay: Status and prospects},}\ }\href {\doibase
  10.1146/annurev-nucl-101918-023407} {\bibfield  {journal} {\bibinfo
  {journal} {Annual Review of Nuclear and Particle Science}\ }\textbf {\bibinfo
  {volume} {69}},\ \bibinfo {pages} {219--251} (\bibinfo {year} {2019})},\
  \Eprint
  {http://arxiv.org/abs/https://doi.org/10.1146/annurev-nucl-101918-023407}
  {https://doi.org/10.1146/annurev-nucl-101918-023407} \BibitemShut {NoStop}%
\bibitem [{\citenamefont {Goeppert-Mayer}(1935)}]{DBD}%
  \BibitemOpen
  \bibfield  {author} {\bibinfo {author} {\bibfnamefont {M.}~\bibnamefont
  {Goeppert-Mayer}},\ }\bibfield  {title} {\enquote {\bibinfo {title} {{Double
  Beta-Disintegration}},}\ }\href {\doibase 10.1103/PhysRev.48.512} {\bibfield
  {journal} {\bibinfo  {journal} {Phys. Rev.}\ }\textbf {\bibinfo {volume}
  {48}},\ \bibinfo {pages} {512--516} (\bibinfo {year} {1935})}\BibitemShut
  {NoStop}%
\bibitem [{\citenamefont {Deppisch}\ \emph {et~al.}(2018)\citenamefont
  {Deppisch}, \citenamefont {Graf}, \citenamefont {Harz},\ and\ \citenamefont
  {Huang}}]{Asym1}%
  \BibitemOpen
  \bibfield  {author} {\bibinfo {author} {\bibfnamefont {F.~F.}\ \bibnamefont
  {Deppisch}}, \bibinfo {author} {\bibfnamefont {L.}~\bibnamefont {Graf}},
  \bibinfo {author} {\bibfnamefont {J.}~\bibnamefont {Harz}}, \ and\ \bibinfo
  {author} {\bibfnamefont {W.-C.}\ \bibnamefont {Huang}},\ }\bibfield  {title}
  {\enquote {\bibinfo {title} {Neutrinoless double beta decay and the baryon
  asymmetry of the universe},}\ }\href {\doibase 10.1103/PhysRevD.98.055029}
  {\bibfield  {journal} {\bibinfo  {journal} {Phys. Rev. D}\ }\textbf {\bibinfo
  {volume} {98}},\ \bibinfo {pages} {055029} (\bibinfo {year}
  {2018})}\BibitemShut {NoStop}%
\bibitem [{\citenamefont {Adams}\ and\ \citenamefont {al.
  (CUORE~collaboration)}(2022)}]{CUORE2022}%
  \BibitemOpen
  \bibfield  {author} {\bibinfo {author} {\bibfnamefont {D.~Q.}\ \bibnamefont
  {Adams}}\ and\ \bibinfo {author} {\bibnamefont {al. (CUORE~collaboration)}},\
  }\bibfield  {title} {\enquote {\bibinfo {title} {{Search for Majorana
  neutrinos exploiting millikelvin cryogenics with CUORE}},}\ }\href {\doibase
  10.1038/s41586-022-04497-4} {\bibfield  {journal} {\bibinfo  {journal}
  {Nature}\ }\textbf {\bibinfo {volume} {604}},\ \bibinfo {pages} {53--58}
  (\bibinfo {year} {2022})}\BibitemShut {NoStop}%
\bibitem [{\citenamefont {Alduino}\ and\ \citenamefont {al.
  (CUORE~collaboration)}(2017)}]{CUOREbckg}%
  \BibitemOpen
  \bibfield  {author} {\bibinfo {author} {\bibfnamefont {C.}~\bibnamefont
  {Alduino}}\ and\ \bibinfo {author} {\bibnamefont {al.
  (CUORE~collaboration)}},\ }\bibfield  {title} {\enquote {\bibinfo {title}
  {{The projected background for the CUORE experiment}},}\ }\href {\doibase
  10.1140/epjc/s10052-017-5080-6} {\bibfield  {journal} {\bibinfo  {journal}
  {{Eur.Phys.J.C}}\ }\textbf {\bibinfo {volume} {77}},\ \bibinfo {pages} {543}
  (\bibinfo {year} {2017})}\BibitemShut {NoStop}%
\bibitem [{\citenamefont {{The CUPID Interest Group}}(2019)}]{CUPIDCDR}%
  \BibitemOpen
  \bibfield  {author} {\bibinfo {author} {\bibnamefont {{The CUPID Interest
  Group}}},\ }\bibfield  {title} {\enquote {\bibinfo {title} {Cupid pre-cdr},}\
  }\href {\doibase 10.48550/arxiv.1907.09376} {\  (\bibinfo {year} {2019}),\
  10.48550/arxiv.1907.09376}\BibitemShut {NoStop}%
\bibitem [{\citenamefont {Poda}(2021)}]{SCINTILLATION}%
  \BibitemOpen
  \bibfield  {author} {\bibinfo {author} {\bibfnamefont {D.}~\bibnamefont
  {Poda}},\ }\bibfield  {title} {\enquote {\bibinfo {title} {Scintillation in
  low-temperature particle detectors},}\ }\href {\doibase
  10.3390/physics3030032} {\bibfield  {journal} {\bibinfo  {journal} {Physics}\
  }\textbf {\bibinfo {volume} {3}},\ \bibinfo {pages} {473--535} (\bibinfo
  {year} {2021})}\BibitemShut {NoStop}%
\bibitem [{\citenamefont {Armengaud}\ and\ \citenamefont {al.
  (CUPID-Mo~collaboration)}(2020)}]{CMO}%
  \BibitemOpen
  \bibfield  {author} {\bibinfo {author} {\bibfnamefont {E.}~\bibnamefont
  {Armengaud}}\ and\ \bibinfo {author} {\bibnamefont {al.
  (CUPID-Mo~collaboration)}},\ }\bibfield  {title} {\enquote {\bibinfo {title}
  {{The CUPID-Mo experiment for neutrinoless double-beta decay: performance and
  prospects}},}\ }\href {\doibase 10.1140/epjc/s10052-019-7578-6} {\bibfield
  {journal} {\bibinfo  {journal} {The European Physical Journal C}\ }\textbf
  {\bibinfo {volume} {80}},\ \bibinfo {pages} {44} (\bibinfo {year}
  {2020})}\BibitemShut {NoStop}%
\bibitem [{\citenamefont {Augier}\ and\ \citenamefont {al.
  (CUPID-Mo~collaboration)}(2022)}]{CUPID_Mo}%
  \BibitemOpen
  \bibfield  {author} {\bibinfo {author} {\bibfnamefont {C.}~\bibnamefont
  {Augier}}\ and\ \bibinfo {author} {\bibnamefont {al.
  (CUPID-Mo~collaboration)}},\ }\bibfield  {title} {\enquote {\bibinfo {title}
  {{Final results on the $0\nu\beta\beta$ decay half-life limit of $^{100}$Mo
  from the CUPID-Mo experiment}},}\ }\href@noop {} {\  (\bibinfo {year}
  {2022})},\ \Eprint {http://arxiv.org/abs/2202.08716} {arXiv:2202.08716
  [nucl-ex]} \BibitemShut {NoStop}%
\bibitem [{\citenamefont {Bellini}\ and\ \citenamefont
  {al.}(2012)}]{Cerenkov_TeO}%
  \BibitemOpen
  \bibfield  {author} {\bibinfo {author} {\bibfnamefont {F.}~\bibnamefont
  {Bellini}}\ and\ \bibinfo {author} {\bibnamefont {al.}},\ }\bibfield  {title}
  {\enquote {\bibinfo {title} {{Measurements of the Cerenkov light emitted by a
  TeO2 crystal}},}\ }\href {\doibase 10.1088/1748-0221/7/11/P11014} {\bibfield
  {journal} {\bibinfo  {journal} {Journal of Instrumentation}\ }\textbf
  {\bibinfo {volume} {7}} (\bibinfo {year} {2012}),\
  10.1088/1748-0221/7/11/P11014}\BibitemShut {NoStop}%
\bibitem [{\citenamefont {Armatol}\ and\ \citenamefont {al.
  (BINGO~collaboration)}(2022)}]{BINGO1}%
  \BibitemOpen
  \bibfield  {author} {\bibinfo {author} {\bibfnamefont {A.}~\bibnamefont
  {Armatol}}\ and\ \bibinfo {author} {\bibnamefont {al.
  (BINGO~collaboration)}},\ }\bibfield  {title} {\enquote {\bibinfo {title}
  {{First cryogenic tests on BINGO innovations}},}\ }\href {\doibase
  10.48550/arxiv.2204.14161} {\  (\bibinfo {year} {2022}),\
  10.48550/arxiv.2204.14161}\BibitemShut {NoStop}%
\bibitem [{\citenamefont {Novati}\ \emph {et~al.}(2019)\citenamefont {Novati},
  \citenamefont {Bergé}, \citenamefont {Dumoulin}, \citenamefont {Giuliani},
  \citenamefont {Mancuso}, \citenamefont {{de Marcillac}}, \citenamefont
  {Marnieros}, \citenamefont {Olivieri}, \citenamefont {Poda}, \citenamefont
  {Tenconi},\ and\ \citenamefont {Zolotarova}}]{NLffect}%
  \BibitemOpen
  \bibfield  {author} {\bibinfo {author} {\bibfnamefont {V.}~\bibnamefont
  {Novati}}, \bibinfo {author} {\bibfnamefont {L.}~\bibnamefont {Bergé}},
  \bibinfo {author} {\bibfnamefont {L.}~\bibnamefont {Dumoulin}}, \bibinfo
  {author} {\bibfnamefont {A.}~\bibnamefont {Giuliani}}, \bibinfo {author}
  {\bibfnamefont {M.}~\bibnamefont {Mancuso}}, \bibinfo {author} {\bibfnamefont
  {P.}~\bibnamefont {{de Marcillac}}}, \bibinfo {author} {\bibfnamefont
  {S.}~\bibnamefont {Marnieros}}, \bibinfo {author} {\bibfnamefont
  {E.}~\bibnamefont {Olivieri}}, \bibinfo {author} {\bibfnamefont
  {D.}~\bibnamefont {Poda}}, \bibinfo {author} {\bibfnamefont {M.}~\bibnamefont
  {Tenconi}}, \ and\ \bibinfo {author} {\bibfnamefont {A.}~\bibnamefont
  {Zolotarova}},\ }\bibfield  {title} {\enquote {\bibinfo {title}
  {Charge-to-heat transducers exploiting the neganov-trofimov-luke effect for
  light detection in rare-event searches},}\ }\href {\doibase
  https://doi.org/10.1016/j.nima.2019.06.044} {\bibfield  {journal} {\bibinfo
  {journal} {Nuclear Instruments and Methods in Physics Research Section A:
  Accelerators, Spectrometers, Detectors and Associated Equipment}\ }\textbf
  {\bibinfo {volume} {940}},\ \bibinfo {pages} {320--327} (\bibinfo {year}
  {2019})}\BibitemShut {NoStop}%
\bibitem [{\citenamefont {Berg\'e}\ and\ \citenamefont {al.}(2018)}]{NLTeO}%
  \BibitemOpen
  \bibfield  {author} {\bibinfo {author} {\bibfnamefont {L.}~\bibnamefont
  {Berg\'e}}\ and\ \bibinfo {author} {\bibnamefont {al.}},\ }\bibfield  {title}
  {\enquote {\bibinfo {title} {{Complete event-by-event
  $\ensuremath{\alpha}/\ensuremath{\gamma}(\ensuremath{\beta})$ separation in a
  full-size ${\mathrm{TeO}}_{2}$ CUORE bolometer by Neganov-Luke-magnified
  light detection }},}\ }\href {\doibase 10.1103/PhysRevC.97.032501} {\bibfield
   {journal} {\bibinfo  {journal} {Phys. Rev. C}\ }\textbf {\bibinfo {volume}
  {97}},\ \bibinfo {pages} {032501} (\bibinfo {year} {2018})}\BibitemShut
  {NoStop}%
\bibitem [{\citenamefont {Haller}\ \emph {et~al.}(1984)\citenamefont {Haller},
  \citenamefont {Palaio}, \citenamefont {Rodder}, \citenamefont {Hansen},\ and\
  \citenamefont {Kreysa}}]{NTDGe}%
  \BibitemOpen
  \bibfield  {author} {\bibinfo {author} {\bibfnamefont {E.~E.}\ \bibnamefont
  {Haller}}, \bibinfo {author} {\bibfnamefont {N.~P.}\ \bibnamefont {Palaio}},
  \bibinfo {author} {\bibfnamefont {M.}~\bibnamefont {Rodder}}, \bibinfo
  {author} {\bibfnamefont {W.~L.}\ \bibnamefont {Hansen}}, \ and\ \bibinfo
  {author} {\bibfnamefont {E.}~\bibnamefont {Kreysa}},\ }\enquote {\bibinfo
  {title} {{NTD Germanium: A Novel Material for Low Temperature Bolometers}},}\
  in\ \href {\doibase 10.1007/978-1-4613-2695-3_2} {\emph {\bibinfo {booktitle}
  {Neutron Transmutation Doping of Semiconductor Materials}}},\ \bibinfo
  {editor} {edited by\ \bibinfo {editor} {\bibfnamefont {R.~D.}\ \bibnamefont
  {Larrabee}}}\ (\bibinfo  {publisher} {Springer US},\ \bibinfo {address}
  {Boston, MA},\ \bibinfo {year} {1984})\ pp.\ \bibinfo {pages}
  {21--36}\BibitemShut {NoStop}%
\bibitem [{\citenamefont {Khalife}(2021)}]{khalife}%
  \BibitemOpen
  \bibfield  {author} {\bibinfo {author} {\bibfnamefont {H.}~\bibnamefont
  {Khalife}},\ }\emph {\bibinfo {title} {{CROSS and CUPID-Mo : future
  strategies and new results in bolometric search for 0$\nu$$\beta$$\beta$}}},\
  \href {https://tel.archives-ouvertes.fr/tel-03168547} {\bibinfo {type}
  {Theses}},\ \bibinfo  {school} {{Universit{\'e} Paris-Saclay}} (\bibinfo
  {year} {2021})\BibitemShut {NoStop}%
\end{thebibliography}%

\end{document}